# Affine Combination of Two Adaptive Sparse Filters for Estimating Large Scale MIMO Channels

Guan Gui and Li Xu
Department of Electronics and Information Systems,
Akita Prefectural University, Akita, Japan
E-mails: {guiguan,xuli}@akita-pu.ac.jp

*Abstract*—Large scale multiple-input multiple-output (MIMO) system is considered one of promising technologies for realizing next-generation wireless communication system (5G) to increasing the degrees of freedom in space and enhancing the link reliability while considerably reducing the transmit power. However, large scale MIMO system design also poses a big challenge to traditional one-dimensional channel estimation techniques due to high complexity and curse of dimensionality problems which are caused by long delay spread as well as large number antenna. Since large scale MIMO channels often exhibit sparse or/and cluster-sparse structure, in this paper, we propose a simple affine combination of adaptive sparse channel estimation method for reducing complexity and exploiting channel sparsity in the large scale MIMO system. First, problem formulation and standard affine combination of adaptive least mean square (LMS) algorithm are introduced. Then we proposed an effective affine combination method with two sparse LMS filters and designed an approximate optimum affine combiner according to stochastic gradient search method as well. Later, to validate the proposed algorithm for estimating large scale MIMO channel, computer simulations are provided to confirm effectiveness of the proposed algorithm which can achieve better estimation performance than the conventional one as well as traditional method.

## I. INTRODUCTION

### A. Background and motivation

Large scale multiple-input multiple-output (MIMO) (see Fig. 1), is considered as one of most promising technologies in the design of future fifth generation (5G) cellular networks [1]. One of main reason is the attractive spectral efficiency up to several tens of bps/Hz which could be achieved by large scale MIMO systems with tens of antennas or even more [2]–[4]. For example, NTT DoCoMo has demonstrated the field experiment of a large scale MIMO system equipped with 12×12 antennas, which approximately achieves the spectral efficiency of 50 bps/Hz with the transmission rate of 4.92 Gbps over a 100 MHz channel bandwidth [5]. To realize large scale MIMO systems, one should solve several key technical challenges as follows: 1) low-complexity signal detection algorithms for practical implementation; 2) proper antenna placement to ensure independent channels; 3) low-complexity precoding algorithms to mitigate the inter-user interference; 4) channel estimation of the high dimensional MIMO channel matrix, etc. [1]–[4]. This paper will focus on low-complexity channel estimation for the large-scale MIMO systems.

In last couple years, several simple adaptive filtering based channel estimation methods, e.g., least mean square (LMS) filter [6], have been proposed for estimating wireless channels. It is well known that the step-size of LMS filter is a critical parameter to balance between convergence speed and steady-state mean square error (MSE) performance. In other words, a faster (slower) convergence speed of LMS filter often yields a higher (lower) steady-state MSE. Indeed, the LMS filter can realize invariable tradeoff with initial empirical step-size, but it is unable to adjust the convergence speed and MSE with the fixed step-size in the updating progress. Hence, an unsuitable step-size either causes performance loss or reduces convergence speed.

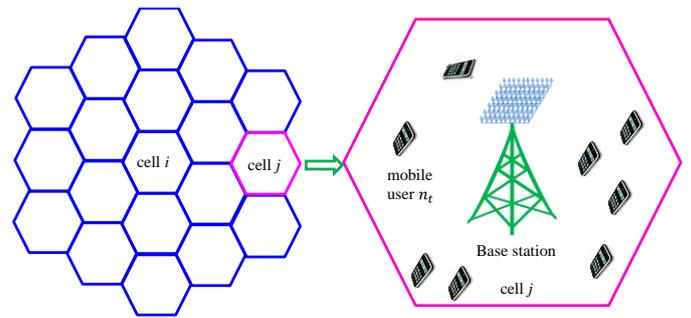

Figure 1. A typical example of large scale MIMO systems.

To deal with this problem of inefficient tradeoff, affine combination of two standard LMS filters (AC-LMS) [7], as shown in Fig. 2, is attracting a lot of attention in the last decades. The basic idea of affine combination methods is to adopt multiple filters with different step-sizes to replace one filter for realizing adaptive tradeoff. Let us take the affine AC-LMS for example. The first filter uses larger step-size than second filter so that the combination filter can achieve a good/fair tradeoff between convergence speed and steady-state MSE performance. If one does not considering any channel structures, AC-LMS [7] would be an effective filter to estimating large scale MIMO channels.

It is well known that wireless channels are molded sparse, containing only a few large coefficients (active) interspered among many negligible ones (inactive), in many scenarios due to broadband transmission in large scale MIMO systems [8]–[10]. However, AC-LMS filter [7] does not consider the sparse structure of finite impulse response (FIR) in unknown systems. Basically, taking advantage of such sparse prior information can improve the identifying performance. Thus, there is a great interest in exploiting the sparse structure information as well as in adjusting step-size to improve the MSE performance while without scarifying convergence speed on sparse channel estimation in large scale MIMO systems.

Motivated by the compressive sensing (CS) [11], [12], Gu and his collaborators proposed an $\ell_0$-norm LMS (L0LMS) filter [13] for estimation sparse channels. However, L0LMS filter adopts only one step-size which cannot tradeoff estimation performance and convergence speed. Based on this background, affine combination of two L0LMS filters has been proposed [14] to estimation single-input single-output (SISO) sparse channels. To the best of our knowledge, no paper has been reported the combined structure of two sparse LMS filters for estimating large scale MIMO channels.

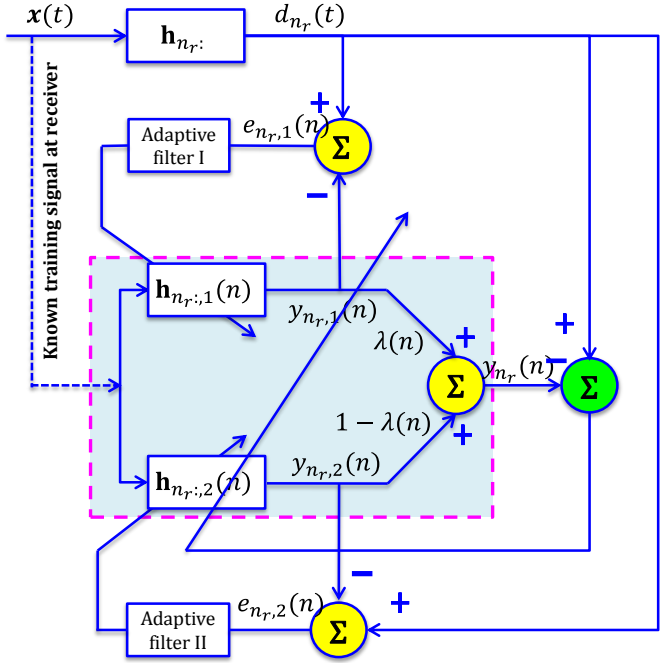

Fig. 2. Affine combination of two adaptive filters for estimating large-scale MIMO channels.

*B. Main contributions*

In this paper, we proposed a large scale MIMO channel estimation method using affine combination of two L0LMS filters (AC-L0LMS). The proposed filter has two properties: 1) Two L0LMS filters with different step-sizes provide a good tradeoff between convergence speed and estimation performance; 2) AC-L0LMS using approximate optimal sparse constraint takes advantage of channel sparsity effectively.

The main contribution of this paper is summarized as follows: $\ell_0$-norm sparse function is introduced to cost function of standard AC-LMS filter and then, sparse AC-LMS filter (i.e., AC-L0LMS filter) is proposed for estimating large scale MIMO channels. Later, several representative experiments are conducted to confirm the effectiveness of our propose methods. In the first experiment, the steady-state MSE performance of the proposed channel estimate is evaluated with the number of nonzero channel taps as parameter. In the second experiment, assuming the constant number of nonzero taps, the average MSE performance of proposed method is evaluated with the step-size ratio as a parameter.

*C. Organizations and notations*

The reminder of the rest paper is organized as follows. Section II reviews the affine combination of two standard LMS filters and problem formulation of large scale MIMO channel estimation. In Section III, we propose affine combination of two sparse LMS filters to estimation large scale MIMO channel without sacrificing convergence speed. In section IV, the simulation results via MSE metric are presented to confirm the effectiveness of proposed method. Concluding remarks are given in Section V.

Throughout the paper, matrices and vectors are represented by boldface upper case letters and boldface lower case letters, respectively; the superscripts $(\cdot)^T$, $(\cdot)^H$, and $(\cdot)^{-1}$ denote the transpose, the Hermitian transpose, and the inverse operators, respectively; $E\{\cdot\}$ denotes the expectation operator; $\|\boldsymbol{x}\|_0$ is the $\ell_0$-norm operator that counts the number of nonzero taps in $\boldsymbol{x}$.

## II. SYSTEM MODEL AND PROBLEM FORMULATION

We consider a large-scale multi-user MIMO-OFDM time-division duplex (TDD) system where the BS equipped with $N_r$ antennas to serve $N_t$ mobile users ($N_r \gg N_t$). Each mobile user is equipped with single-antenna. In addition, we assume ideal orthogonal pilot designs in all users and hence there is no pilot contamination. At time index $t$, $n_t$-th user frequency domain transmit signal vector $\bar{\boldsymbol{x}}_{n_t}(t) = [\bar{x}_{n_t}(t,0), \cdots, \bar{x}_{n_t}(t, N-1)]^T$, $n_t = 1, 2, \cdots, N_t$ is fed to inverse discrete Fourier transform (IDFT), where $N$ is the number of subcarriers. Assume that the transmit power is $E\{\|\bar{\boldsymbol{x}}_{n_t}(t)\|_2^2\} = NE_0$, where $E_0$ denotes unit power. The resultant vector $\boldsymbol{x}_{n_t}(t) = \boldsymbol{F}^H \bar{\boldsymbol{x}}_{n_t}(t)$ is padded with cyclic prefix (CP) of length $L_{cp} \geq N-1$ to avoid inter-block interference (IBI), where $\boldsymbol{F}$ is a $N \times N$ DFT matrix with entries $[\boldsymbol{F}]_{cq} = 1/\sqrt{N} e^{-j2\pi cq/N}$, $c,q = 0,1,...,N-1$. The time domain signal is transmitted through length $N$ channel and received by multiple antennas at the receiver. After CP removal, the signal vector received by the $n_r$-th antenna at time $t$ is written as $y_{n_r}$. Then, the ideal received signal vector $\boldsymbol{d}(t) = [d_1(t), d_2(t), \cdots, d_{N_r}(t)]^T$ and input signal $\boldsymbol{x}(t)$ are related by

$$\boldsymbol{d}(t) = \boldsymbol{H}\boldsymbol{x}(t) + \boldsymbol{z}(t), \tag{1}$$

where the MIMO channel matrix $\boldsymbol{H}$ can be written as

$$\boldsymbol{H} = \begin{bmatrix} \boldsymbol{h}_{11}^T & \boldsymbol{h}_{12}^T & \cdots & \boldsymbol{h}_{1N_t}^T \\ \boldsymbol{h}_{21}^T & \boldsymbol{h}_{22}^T & \cdots & \boldsymbol{h}_{2N_t}^T \\ \vdots & \vdots & \ddots & \vdots \\ \boldsymbol{h}_{N_r 1}^T & \boldsymbol{h}_{N_r 2}^T & \cdots & \boldsymbol{h}_{N_r N_t}^T \end{bmatrix} = \begin{bmatrix} \boldsymbol{h}_{1:}^T \\ \boldsymbol{h}_{2:}^T \\ \vdots \\ \boldsymbol{h}_{N_r:}^T \end{bmatrix}, \tag{2}$$

Note that $NN_t$ is effective filtering memory length of each MISO channel vector $\boldsymbol{h}_{n_r:}$ between $n_r$-antenna and all mobile

user. As shows in Fig. 2, the ideal received signal at $n_r$-th antenna is

$$d_{n_r}(t) = \sum_{n_t=1}^{N_t} \boldsymbol{h}_{n_r,n_t}^T \boldsymbol{x}_{n_t}(t) + z_{n_r}(t) = \boldsymbol{h}_{n_r,:}^T \boldsymbol{x}(t) + z_{n_r}(t), \quad (3)$$

where $\boldsymbol{h}_{n_r,:} = [\boldsymbol{h}_{n_r,1}^T, \cdots, \boldsymbol{h}_{n_r,n_t}^T, \cdots, \boldsymbol{h}_{n_r,N_t}^T]^T \in \mathbb{C}^{N_t N}$, $n_r = 1, \cdots, N_r$ is an MISO channel vector which consists of $N_t$ single-input single-output (SISO) sub-channels $\boldsymbol{h}_{n_r,n_t}$; $z_{n_r}(n)$ denotes an additive Gaussian noise variable with distribution $\mathcal{CN}(0, \sigma_n^2)$. We assume that the sub-channel $\boldsymbol{h}_{n_r,n_t}$ is only supported by $K$ ($K \ll N$) nonzero channel taps whose positions are randomly determined. Sparse channel model assumption is suitable due to the fact that broadband signal transmission is adopted in large-scale MIMO systems [15]. Hereby, at the $n_r$-th receive antenna at BS, the corresponding $i$-th filter estimation error $e_{n_r,:i}(n)$, for $n_r = 1, 2, \cdots, N_r$ at time $t$ can be written as:

$$\begin{aligned} e_{n_r,i}(n) &= d_{n_r}(t) - y_{n_r,i}(n) \\ &= d_{n_r}(t) - \boldsymbol{h}_{n_r,:i}^T(n)\boldsymbol{x}(t), \ i=1,2, \end{aligned} \quad (4)$$

where $\boldsymbol{h}_{n_r,:i}(n)$ denotes the $n_r$-th channel estimate and $y_{n_r,i}(n)$ is the received signal from $i$-th filter. By collecting all of the error signals $e_{n_r,i}(n)$, $n_r = 1, 2, \cdots, N_r$, Eq. (4) can be also written as matrix-vector form:

$$\begin{aligned} \boldsymbol{e}_i(n) &= [e_{1,i}(n), e_{2,i}(n), \ldots, e_{N_r,i}(n)]^T \\ &= \boldsymbol{d} - \boldsymbol{y}_i(n) = \boldsymbol{d} - \boldsymbol{H}_i(n)\boldsymbol{x}(t), \ i=1,2, \end{aligned} \quad (5)$$

where $\boldsymbol{y}_i(n) = [y_{1,i}(n), \cdots, y_{N_r,i}(n)]^T$ denotes estimate of the output signal; $\boldsymbol{H}_i(n)$ is the $n$-th adaptive large MIMO channel estimate $\boldsymbol{H}_i$ with $i$-th filter. According to Eq. (5), large scale MIMO adaptive channel estimation problem is equivalent to estimate $N_r$ individual MISO channel vectors $\boldsymbol{h}_{n_r}$. Hence, MISO channel can be estimated by standard LMS filter [6] which cost function is

$$L_{n_r,i}(n) = (1/2)e_{n_r,i}^2(n), \ i=1,2, \quad (6)$$

for $n_r = 1, 2, \cdots, N_r$. Its updating equation can be derived as

$$\begin{aligned} \boldsymbol{h}_{n_r,:i}(n+1) &= \boldsymbol{h}_{n_r,:i}(n) - \mu_i \frac{\partial L_{n_r,i}(n)}{\partial \boldsymbol{h}_{n_r,:i}(n)} \\ &= \boldsymbol{h}_{n_r,:i}(n) + \mu_i \boldsymbol{x}(t)e_{n_r,i}(n), \ i=1,2, \end{aligned} \quad (7)$$

where $\mu_i \in (0, \gamma_{\max}^{-1})$ is the step size of LMS gradient descend and $\gamma_{\max}$ is the maximum eigenvalue of the $NN_t \times NN_t$ covariance matrix, which is calculated as $\boldsymbol{R} = E\{\boldsymbol{x}(n)\boldsymbol{x}^H(n)\}$. Without loss of generality, we assume $\mu_2 = \delta\mu_1$ ($0 < \delta < 1$) so that $\boldsymbol{h}_{n_r,:1}(n)$ achieves faster convergence speed than $\boldsymbol{h}_{n_r,:2}(n)$. Notice that the steady-state MSE performance of the $\boldsymbol{h}_{n_r,:2}(n)$ is better than $\boldsymbol{h}_{n_r,:1}(n)$. Also, assuming both $\boldsymbol{h}_{n_r,:1}(n)$ and $\boldsymbol{h}_{n_r,:2}(n)$ are coupled deterministically and statistically through input signal vector $\boldsymbol{x}(t)$ and additive noise variable $z_{n_r}(t)$. Assuming two individual output signal $y_{n_r,i}(n) = \boldsymbol{h}_{n_r,:i}^T(n)\boldsymbol{x}(t)$, $i=1,2$ are independent. According to

Fig. 2, the $n_r$-th receive signal $y_{n_r}(n)$ of the affine combination of the two LMS filters is given by

$$\begin{aligned} y_{n_r}(n) &= \lambda(n)y_{n_r,1}(n) + (1-\lambda(n))y_{n_r,2}(n) \\ &= \lambda(n)\boldsymbol{h}_{n_r,:1}^T(n)\boldsymbol{x}(t) + (1-\lambda(n))\boldsymbol{h}_{n_r,:2}^T(n)\boldsymbol{x}(t) \\ &= \left[\lambda(n)\left(\boldsymbol{h}_{n_r,:1}(n) - \boldsymbol{h}_{n_r,:2}(n)\right) + \boldsymbol{h}_{n_r,:2}(n)\right]^T \boldsymbol{x}(t) \\ &= \left[\lambda(n)\boldsymbol{h}_{n_r,:12}(n) + \boldsymbol{h}_{n_r,:2}(n)\right]^T \boldsymbol{x}(t), \end{aligned} \quad (8)$$

where $\boldsymbol{h}_{n_r,:12}(n) = \boldsymbol{h}_{n_r,:1}(n) - \boldsymbol{h}_{n_r,:2}(n)$ is a differential filter and $\lambda(n)$ is a affine combination parameter to decide final system identification error. In Eq. (8), one can find that $y_{n_r}(n)$ can be considered as a affine combination of filter $\boldsymbol{h}_{n_r,:2}(n)$ and a weighted differential filter $\lambda(n)\boldsymbol{h}_{n_r,:12}(n)$. Hence, Eq. (8) implies an equivalent filter as:

$$\boldsymbol{h}_{n_r,:eq} = \lambda(n)\boldsymbol{h}_{n_r,:12}(n) + \boldsymbol{h}_{n_r,:2}(n), \quad (9)$$

According to (4) and (9), the overall system error is given by

$$\begin{aligned} e_{n_r}(n) &= d_{n_r}(t) - y_{n_r}(n) \\ &= \left[\boldsymbol{h}_{n_r,:o2}(n) - \lambda(n)\boldsymbol{h}_{n_r,:12}(n)\right]^T \boldsymbol{x}(t) + z(t), \end{aligned} \quad (10)$$

where $\boldsymbol{h}_{n_r,:o2}(n) = \boldsymbol{h}_{n_r,:} - \boldsymbol{h}_{n_r,:2}(n)$ is regarded as a differential filter as well. This setup generalizes the combination of adaptive filter outputs, and hence it can be used to study the properties of the optimal combination. In [7], the authors proposed the optimal affine combiner:

$$\lambda_o(n) = \frac{\boldsymbol{h}_{n_r,:o2}^H(n)\boldsymbol{R}_{xx}\boldsymbol{h}_{n_r,:12}(n)}{\boldsymbol{h}_{n_r,:12}^H(n)\boldsymbol{R}_{xx}\boldsymbol{h}_{n_r,:12}(n)}, \quad (11)$$

which is the expectation for the optimum $\lambda(n)$ as a function of the unknown differential channel vector $\boldsymbol{h}_{n_r,:12}$, where $\boldsymbol{R}_{xx} = E\{\boldsymbol{x}(t)\boldsymbol{x}^T(t) | (\boldsymbol{h}_{n_r,:2}(n), \boldsymbol{h}_{n_r,:12}(n))\}$ denotes the input conditional autocorrelation matrix. It is easy to find that the optimal affine combiner is based on prior knowledge of the unknown system FIR $\boldsymbol{h}_{n_r,:12}$. However, it cannot be utilized in practical channel estimation. Using stochastic gradient search method, suboptimal affine combiner $\lambda_s(n)$ [7] was proposed as follows

$$\lambda_s(n+1) = \lambda_s(n) + \mu_\lambda \left[d_{n_r}(t) - \tilde{\boldsymbol{h}}_{n_r,:12}^T(n)\boldsymbol{x}(t)\right]\boldsymbol{h}_{n_r,:12}^T(n)\boldsymbol{x}(t), \quad (12)$$

where $\tilde{\boldsymbol{h}}_{n_r,:12}(n) = \lambda_s(n)\boldsymbol{h}_{n_r,:1}(n) + (1-\lambda_s(n))\boldsymbol{h}_{n_r,:2}(n)$ is a $n$-th updating affine combination of two estimator vectors (i.e., $\boldsymbol{h}_{n_r,:1}(n)$ and $\boldsymbol{h}_{n_r,:2}(n)$) and $\mu_\lambda$ is empirical parameter for tracking the adaptation of $\boldsymbol{h}_{n_r,:1}(n)$ as well as $\boldsymbol{h}_{n_r,:2}(n)$.

III. PROPOSED METHOD FOR ESTIMATING LARGE SCALE MIMO CHANNEL

Since the affine combination of two standard LMS filters neglects the inherent system sparsity, it often causes performance loss. Unlike the traditional method, we propose an affine combination of two L0LMS filters to exploit the large scale MIMO channel sparsity, with two individual cost functions [13],

$$G_{n_r,i}(n) = (1/2)e_{n_r,i}^2(n) + \beta_i \|\boldsymbol{h}_{n_r,:,i}\|_0, \quad i=1,2, \quad (13)$$

where $\beta_i$ is a positive regularization parameter for tradeoff between estimation error term and sparsity of channel estimate. It is well known that solving the $\|\boldsymbol{h}_{n_r,:,i}\|_0$ in Eq. (13) is a (non-deterministic polynomial-time) NP-hard problem [11]. To deal with this problem, it could be approximated by a continuous function

$$\|\boldsymbol{h}_{n_r,:,i}\|_0 = \sum_{l=0}^{NN_t}(1 - e^{-\alpha|h_{n_r,:,i}^l|}), \quad i=1,2, \quad (14)$$

According to (14), cost function of the $i$-th L0LMS filter can be modified as

$$G_{n_r,i}(n) = (1/2)e_{n_r,i}^2(n) + \beta_i \sum_{l=0}^{NN_t}(1 - e^{-\alpha|h_{n_r,:,i}^l|}), \quad i=1,2. \quad (15)$$

Then, with different step-size $\mu_i$, the $(n+1)$-th update sparse channel estimate is derived as

$$\begin{aligned}\boldsymbol{h}_{n_r,:,i}(n+1) &= \boldsymbol{h}_{n_r,:,i}(n) + \mu_i \frac{\partial G_{n_r,i}(n)}{\partial \boldsymbol{h}_{n_r,:,i}(n)} \\ &= \boldsymbol{h}_{n_r,:,i}(n) + \mu_i e_{n_r,i}(n)\boldsymbol{x}(t) \\ &\quad - \alpha\mu_i\beta_i \operatorname{sgn}\{\boldsymbol{h}_{n_r,:,i}(n)\}e^{-\alpha|\boldsymbol{h}_{n_r,:,i}(n)|},\end{aligned} \quad (16)$$

for $i=1,2$. However, the exponential function $e^{-\alpha|\boldsymbol{h}_{n_r,:,i}(n)|}$ in Eq. (16) still causes high computational complexity. To reduce the high complexity, the first order Taylor series expansion of exponential function is taken into consideration as

$$e^{-\alpha|\boldsymbol{h}_{n_r,:,i}(n)|} \approx \begin{cases} 1 - \alpha|h_{n_r,:,i}^l(n)|, & \text{when } |h_{n_r,:,i}^l(n)| \le 1/\alpha \\ 0, & \text{others.}\end{cases} \quad (17)$$

It is worth mentioning that the positive parameter $\alpha$ controls the channel sparseness and estimation performance. Though the L0LMS can exploit channel sparsity on adaptive channel estimation, unsuitable threshold parameter $\alpha$ will cause overall identification performance degradation. In this paper, we adopted $\alpha=10$ which is also suggested as in [16]. According to above analysis, the modified update equation of L0LMS can be rewritten as

$$\boldsymbol{h}_{n_r,:,i}(n+1) = \boldsymbol{h}_{n_r,:,i}(n) + \mu_i e_{n_r,i}(n)\boldsymbol{x}(t) - \alpha\mu_i\beta_i\mathcal{S}\{\boldsymbol{h}_{n_r,:,i}(n)\}, \quad (18)$$

where the $\ell_0$-norm sparse penalty approximation function $\mathcal{S}\{\boldsymbol{h}_{n_r,:,i}(n)\}$ is defined as

$$\mathcal{S}\{\boldsymbol{h}_{n_r,:,i}(n)\} = \begin{cases} 2\alpha^2 h_{n_r,:,i}^l(n) - 2\alpha\operatorname{sgn}\{h_{n_r,:,i}^l(n)\}, & \text{if } |h_{n_r,:,i}^l(n)| \le 1/\alpha \\ 0, & \text{others}\end{cases} \quad (19)$$

Analogy to Eq. (12), suboptimal affine combiner $\lambda_s(n)$ of two L0LMS filters can also be given by

$$\lambda_s(n+1) = \lambda_s(n) + \mu_\lambda \left[d_{n_r}(t) - \tilde{\boldsymbol{h}}_{n_r,:,12}^T(n)\boldsymbol{x}(t)\right]\boldsymbol{h}_{n_r,:,12}^T(n)\boldsymbol{x}(t), \quad (20)$$

where $\tilde{\boldsymbol{h}}_{n_r,:,12}(n) = \tilde{\lambda}_s(n)\boldsymbol{h}_{n_r,:,1}(n) + (1-\tilde{\lambda}_s(n))\boldsymbol{h}_{n_r,:,2}(n)$ and $\mu_\lambda$ is parameter for tracking the adaptation of sparse estimates, i.e., $\boldsymbol{h}_{n_r,:,1}(n)$ and $\boldsymbol{h}_{n_r,:,2}(n)$. By exploiting channel sparsity, output signal $\tilde{y}_{n_r}(n)$ of affine combination of the two L0LMS filters is given by

$$\begin{aligned}\tilde{y}_{n_r}(n) &= \lambda_s(n)\tilde{y}_{n_r,1}(n) + (1-\lambda_s(n))\tilde{y}_{n_r,2}(n) \\ &= \left[\lambda_s(n)\boldsymbol{h}_{n_r,:,12}(n) + \boldsymbol{h}_{n_r,:,2}(n)\right]^T \boldsymbol{x}(t).\end{aligned} \quad (21)$$

Finally, the overall system error $e_{n_r}(n)$ can be computed as

$$\begin{aligned}e_{n_r}(n) &= d_{n_r}(t) - y_{n_r}(n) \\ &= \left[\boldsymbol{h}_{n_r,:,o2}(n) - \lambda_s(n)\boldsymbol{h}_{n_r,:,12}(n)\right]^T \boldsymbol{x}(t) + z(t),\end{aligned} \quad (22)$$

where $\boldsymbol{h}_{n_r,:,o2}(n) = \boldsymbol{h}_{n_r,:} - \boldsymbol{h}_{n_r,:,2}(n)$ denotes a differential filter between real and sparse FIR vector.

## IV. COMPUTER SIMULATIONS AND DISCUSSIONS

To validate effectiveness of the proposed method, we evaluate the average MSE performance which is defined as

$$\begin{aligned}MSE\{\Delta \boldsymbol{H}(n)\} &= E\left\{\|\boldsymbol{H} - \boldsymbol{H}(n)\|_2^2\right\} \\ &= \sum_{n_r=1}^{N_r} E\left\{\|\boldsymbol{h}_{n_r,:} - \boldsymbol{h}_{n_r,:}(n)\|_2^2\right\}.\end{aligned} \quad (23)$$

The results are averaged over 1000 independent Monte-Carlo runs. The signal-to-noise ratio (SNR) is defined as $E_s/\sigma_n^2$, where $E_s$ is the received power of the input signal. Detailed computer simulation parameters are listed in Table. I.

TAB. I. SIMULATION PARAMETERS.

| parameters | values |
|---|---|
| # of antenna at BS | 24 |
| # of mobile users in a celll | 8 |
| length of FIR sub-channel | $N=16$ |
| # of nonzero coefficients | $K=1$ and $4$ |
| positions of nonzero coefficients | Random allocation |
| distribution of FIR coefficient | random Gaussian $\mathcal{CN}(0,1)$ |
| training signal | Pseudorandom noise sequence |
| SNR | {10dB, 20dB} |
| step-size of filter I and filter II | $\mu_1 = 1/N + \gamma$ and $\mu_2 = \delta\mu_1$ |
| controlling the ratio $\delta = \mu_1/\mu_2$ | $\delta \in (0,1)$ |
| controlling $\mu_1$ of the filter I | $\gamma = 4$ |
| tracking adaptation of $\boldsymbol{h}_{n_r,i}$ ($i=1,2$) in $\lambda_s(n)$ | $\mu_\lambda = 1$ |
| parameters for $\ell_0$-norm sparse penalty | $\beta = 0.02\sigma_n^2$ and $\alpha = 10$ |

In the first example, considering two different step-sizes, i.e., $\mu_1 = 1/(NN_t + \gamma)$ and $\mu_2 = 0.5\mu_1$, steady-state MSE performance of proposed estimate is evaluated in the case of $K=1$ and 4, respectively. To verify the effectiveness of the proposed method, we compare it with three previous methods, i.e., LMS [6], L0LMS [13] and AC-LMS [7] as show in Figs.

3-6. For a fair comparison of these methods, same regularization parameter is adopted for L0LMS and AC-L0LMS algorithms, i.e., $\beta = 0.02\sigma_n^2$, which is also recommended by [17][18]. As Figs. 3~6 show, our proposed method can achieve lower MSE performance without reducing the convergence speed. Note that choosing smaller (bigger) $\delta$ can achieve lower (higher) MSE and faster (slower) convergence speed. In addition, MSE performance of the proposed method also depends on the channel sparseness in real large scale MIMO systems. For sparser channel, the proposed method can achieve much lower MSE performance by comparing MSE curves of proposed method in Fig. 3 ($K=1$) and Fig. 4 ($K=4$) in the case of SNR=10dB. Hence, the proposed method should choose different empirical parameters (e.g., $\delta$ and $\beta$) to meet actual requirements of the wireless communications. Due to space limitations, the optimal parameters selection methods and performance analysis will be discussed in our coming journal paper.

In the second example, the proposed channel estimate is also evaluated with respect to different $\delta$ ratio which controls the filter II. Note that the step-size of filter I was fixed $\mu_1 = 1/(NN_t + \gamma)$ and the step-size of filter II was set as $\mu_2 = \delta\mu_1$, where $\delta \in (0,1]$. It is well known that LMS filter using larger (smaller) step-size obtains lower (higher) steady-state MSE performance with faster (slower) convergence speed. Since the step-size ratio $\delta$ is a critical parameter which controls the convergence speed and steady-state MSE performance of the proposed method. In this experiment, five ratios $\delta \in \{0,1,0.3,0.5,0.7,0.9\}$ are adopted for comparisons as shown in Figs. 7 and 8. In the two figures, $\delta = 0.1$ and $\delta = 0.9$ are utilized in a standard AC-LMS filter as for performance benchmarks. We observe that ratio of two step-sizes can balance between estimation performance and convergence speed. For example, larger ratio ($\delta = 0.9$) realizes faster convergence speed but obtains worse MSE performance. In turn, smaller ratio ($\delta = 0.1$) can obtain the better estimation performance while scarifies convergence speed. Hence, simulation results in Figs. 7~8 indicate that suitable step-size ratio is selected as either $\delta = 0.5$ or $\delta = 0.7$.

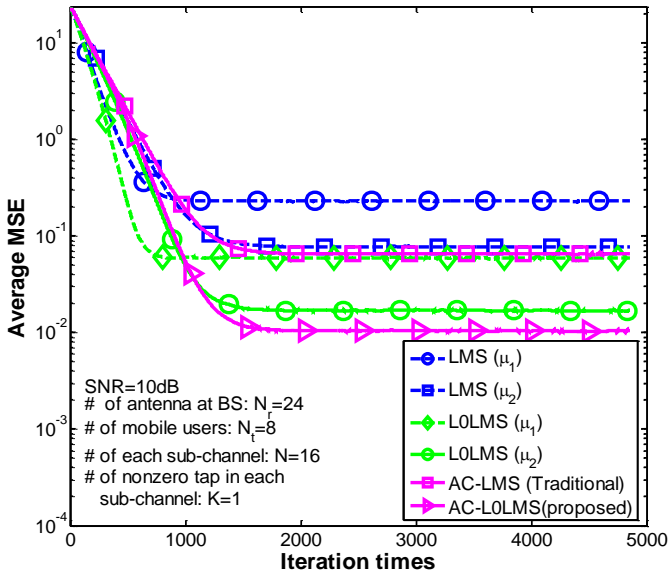

Fig. 3. Average MSE performance comparisons: SNR=10dB and $K$=1.

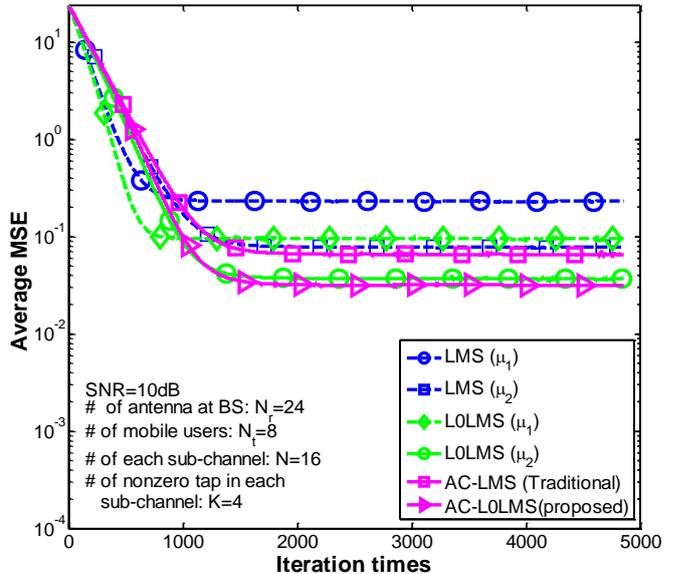

Fig. 4. Average MSE performance comparisons: SNR=10dB and $K$=4.

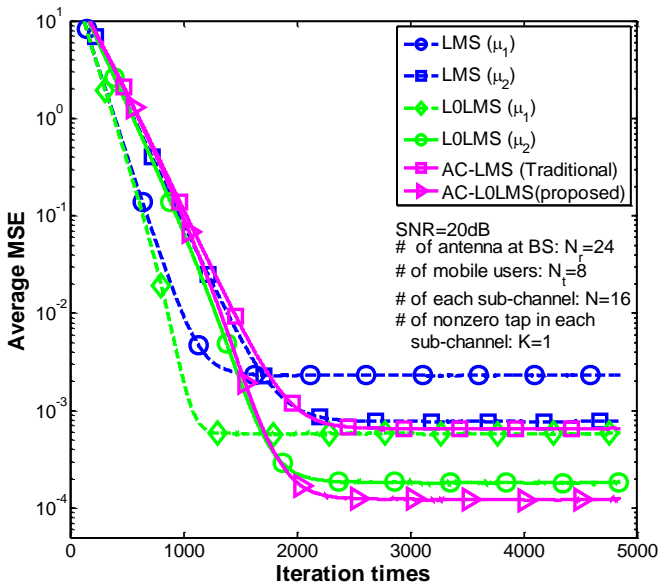

Fig. 5. Average MSE performance comparisons: SNR=20dB and $K$=1.

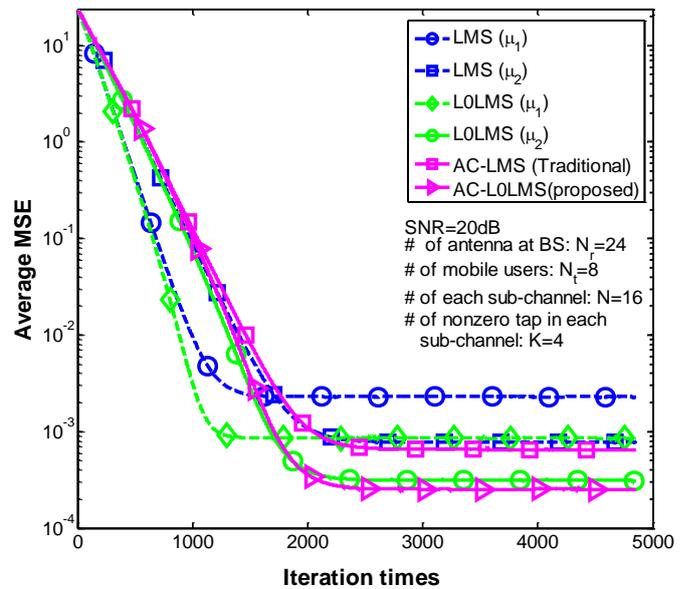

Fig. 6. Average MSE performance comparisons: SNR=20dB and $K$=4.

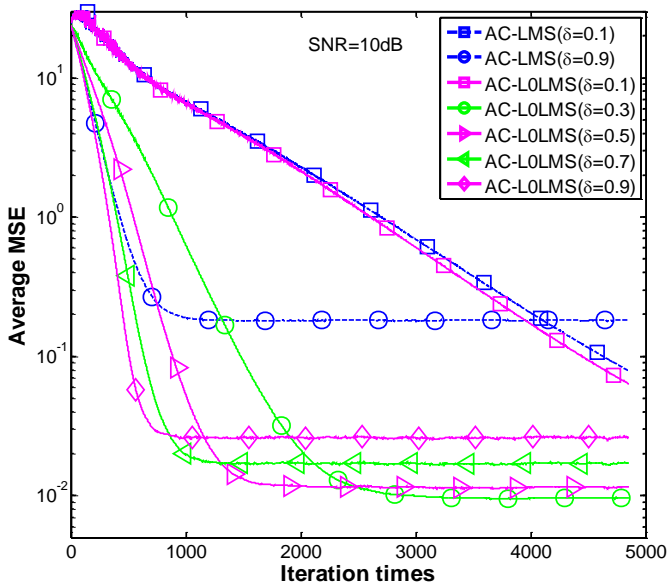

Fig. 6. Average MSE performance comparisons in different $\delta$ (SNR=10dB).

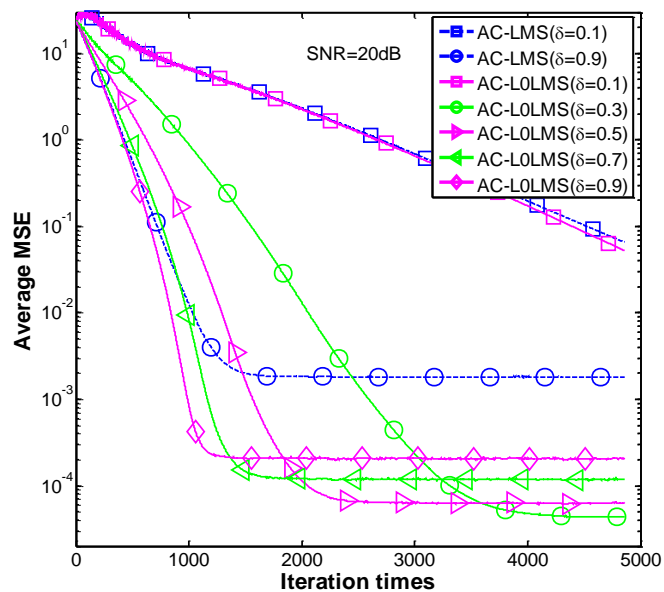

Fig. 7. Average MSE performance comparisons in different $\delta$ (SNR=20dB).

## V. CONCLUDING REMARKS

Traditional channel estimation methods used to apply only one sparse LMS filter with an invariable step-size which cannot balance well between steady-state MSE performance and convergence speed. Hence, they are vulnerable to either performance loss or convergence speed deceleration. In other words, they cannot simultaneously achieve fast convergence speed and high steady-state MSE performance. Unlike these traditional methods, in this paper, we proposed an affine combination of two sparse LMS filters which can achieve fast convergence and high steady-state MSE performance to improve estimation performance. First, problem formulation and standard affine combination of LMS filters were introduced. Then, $\ell_0$-norm sparse constraint function based affine combination of two sparse LMS filters for estimating large scale MIMO channels was presented. Channel estimate performance depends on which affine combiner to choose. The approximate optimum affine combiner was adopted for the proposed filter according to stochastic gradient search method. Later, to verify the effectiveness of the proposed method, selected simulations were provided to confirm the effectiveness of the proposed method which can achieve better estimation performance than the conventional one and standard affine combination of LMS filters.

In future work, we will follow this work from two aspects. In the first place, theoretical analysis of the proposed algorithm (AC-L0LMS) is lack. Based on compressive sensing [11], [12] and our previous works [19], [20], comprehensive theoretical analysis will be studied. In the second place, pilot contamination in multi-cell large scale MIMO system is unavoidable due to the non-orthogonal uplink training sequence [2], [21]. In this paper, channel estimation problem was only considered in single-cell but it unsuitable applies directly in realistic communication systems. In future work, we will study a more flexible method for estimating large scale channels as well mitigation of pilot contaminations.